# Spectroscopic observations of novae V1065 CEN and V1280 SCO using 45 cm cassegrain telescope at Arthur C Clarke Institute

S. Gunasekera[1], J. Adassuriya[1], I. Madagangoda[1]
K. Werellapatha[2], K.P.S.C Jayaratne[2]
[1] Arthur C Clarke Institute for Modern Technologies, Katubedda, Moratuwa
[2] Department of Physics, University of Colombo, Colombo 03

**ABSTRACT**

The spectroscopic observations of two novae namely V1065 CEN and V1280 SCO were made by 45 cm Cassegrain telescope in high resolution ($\lambda/\Delta\lambda=22000$) at H$\alpha$ (6563 $^o$A) region. V1065 CEN is He/N-type spectra which characterize a broad (Gaussian FWHM 49 $^o$A), saddle shaped and asymmetric H$\alpha$ emission line without prominent P-Cyg absorption component. Completely different H$\alpha$ profile of V1280 SCO shows prominent P-Cyg absorption and narrow emission line (Gaussian FWHM 26 $^o$A) which can be classified as Fe II type nova. The expansion velocities of these two systems measured from the minima of the P-Cyg profiles are close to 2300 km/s for V1065 CEN, and 716 km/s for V1280 SCO.

Based on the photometric analysis, the Nova V1065 CEN can be classified as fast ($11<t_2<25$) nova. The derived absolute magnitudes at maximum for nova V1065 CEN to be $M_{o,V} = -7.58\pm0.18$ and $M_{o,B}= -7.75\pm0.25$ correspond to a distance $8.51\pm0.33$ kpc. The parameters $t_{2V}$=12 days and $t_{3V}$=14 days of nova V1280 SCO determine that the nova is in between very fast and fast nova. The mean absolute magnitude at maximum is calculated to be $M_{o,V}=-8.7\pm0.1$ and the estimated distance to the nova V1280 SCO is $3.2\pm0.2$ kpc.

## 1. INTRODUCTION

Spectroscopic observation of two novae had done by 45 cm Cassegrain telescope at Arthur C Clarke observatory namely nova Centauri 2007 (V1065 CEN) and nova Scorpii 2007 (V1280 SCO) during the period at 31st January to 20th Feb 2007.

V1065 CEN was discovered by W. Liller and Vina del Mar [1] on January 23.35 UT, at the coordinates RA$_{2000}$ = $11^h$ $43.2^m$, Dec$_{2000}$= $-58^0$ 03' two days before the maximum. According IAU circular 8800, apparent magnitude on the day of discovery is $m_v = 8.7$ and the spectra obtained on 26.3 January show H-alpha line has an equivalent with of about 93.0 nm and there seems to be P-Cyg structure near the Na Doublet lines. The low resolution spectra taken by L. Andrew Helton indicated strong (Ne II) emissions at the early stage at the evolution and confirmed the nova belongs to O-Ne-Mg class.

The second target V1280 SCO was discovered by H. Yamaoka & Yuji Nakamura [2] on Feb 4.86 UT 2007 at position RA$_{2000}$= $16^h$ $57^m$ $41^s$, Dec$_{2000}$ = $-32^0$ 20' 34". The maximum apparent magnitude $m_v$= 4.0 and the magnitude at the first observation $m_v$~ 8.9 suggests that variable star is a classical nova observed few days before the maximum.

The spectroscopic observation of H-α region was observed using 45cm Cassegrain telescope at Arthur C Clarke institute. The H-α profile of V1065 CEN were obtained 6, 15, and 26 days after maximum. Highly structured broad emission peaks were observed with the expansion velocity around 2000 km/s. The H-α Profiles of V1280 SCO were obtained 4 days after maximum. The nova V1280 SCO is too faint to be observed after the first observation and could not be observed further. The photometry analysis has been done for both novae using the light curve data taken from AAVSO web site. Both V and B band light curve data were available for V1065 CEN and for nova V1280 SCO, only visual magnitudes were available.

The main aims of this paper to present the H-α line profile of both novae at different stages and to discuss the meaning of the observed spectra. In addition we also estimate the absolute magnitude (M) and hence the distances to the two novae given by the characteristic light curve parameters $t_2$ and $t_3$.

## 2. SPECTROSCOPIC OBSERVATIONS

The spectroscopic observations were carried out with the spectrograph attached to the 45 cm Cassegrain telescope with f/12 at Arthur C Clarke Institute, Sri Lanka. Three high resolution ($\lambda/\Delta\lambda=22000$) profiles were obtain for V1065 CEN, 6, 15 and 20 days after maximum and two plates of the same resolution were obtained for V1280 SCO 4 days after maximum. The dispersion element of 1200 lines/mm reflective grating is used at first order.

The detector was 756 pixels CCD along the dispersion direction cooled at -6 $^{o}$C. On the CCD image the spectral dispersion of 0.31 $^{o}$A per pixel at wavelength 6563 $^{o}$A and a spectral coverage of 225 $^{o}$A were obtained. The exposure time was 20 minutes for V1065 CEN and was 10 minutes for V1280 SCO. Fe-Ne emission lamp was used for wavelength calibration of the spectra.

The preliminary data reduction of the CCD frames such as dark substation was done using the CCDOPS software. The main steps of the data reduction scheme, aperture extraction, wavelength calibration, normalization to continuum and conversion to the heliocentric velocity scale were done by using IRAF (Image Reduction and Analysis Facility).

### 2.1 Spectroscopic Observations of V1065 CEN

The high resolution spectra in Hα (6563 $^{o}$A) region are plotted in Fig. 1 for nova V1065 CEN in three different days after outburst. Very broad emission lines were observed (v = 2000 km/s) in three velocity profiles and invariant with the time even after 26 days from the maximum. The spectra are double peaked with timely varying blue and red components. Initially the blue peak (5.398) is higher than the red peak (5.213) but after 26 days this has been changed red to 5.148 and blue to 4.855. The velocity profiles of these two peaks are behaved in the same way. Initially (6 days after maximum) the blue peak shows -364 km/s (Fig. 1a) and later it was drop to -316 km/s (Fig. 1c). The red peak is moving further away from the line center having the velocities 179 km/s at the first observation (Fig. 1a) and 275 km/s at 26 days after maximum. There is no prominent P-Cyg structure in three spectra except a possible absorption at -944 km/s.

The time evolution spectra of V1065 CEN show the early Hα profile is highly structured with emission spikes at -1171 km/s, -608 km/s, -371 km/s, 572 km/s and 820 km/s. These emission knots were decline with the time in the plate **c,** 26 days after maximum.

## 2.2 Spectroscopic Observations of V1280 SCO

The second nova was also observed in the same spectral region Hα with the same resolution (Fig. 2). Both spectra were taken on the same day Feb 20.9, 4 days after maximum. The two profiles are in 20 minutes time gap. The nova is too faint to be observed and could not make any observation after this with the limiting magnitude (m≈12) of the telescope. The spectra show this structure is completely different from the previous observation. Especially the line spectra have strong P-Cyg profile with less broadened single peaked with few emission spikes. The P-Cyg absorption trough was blue shifted by -716 km/s and the blue peak was observed at -138 km/s from the profile center and moves towards the profile center with the time as observed in V1065 CEN. The equivalent widths for the emission components of two profiles taken at 10 minutes interval were estimated using IRAF which are 38 °A and 40 °A respectively.

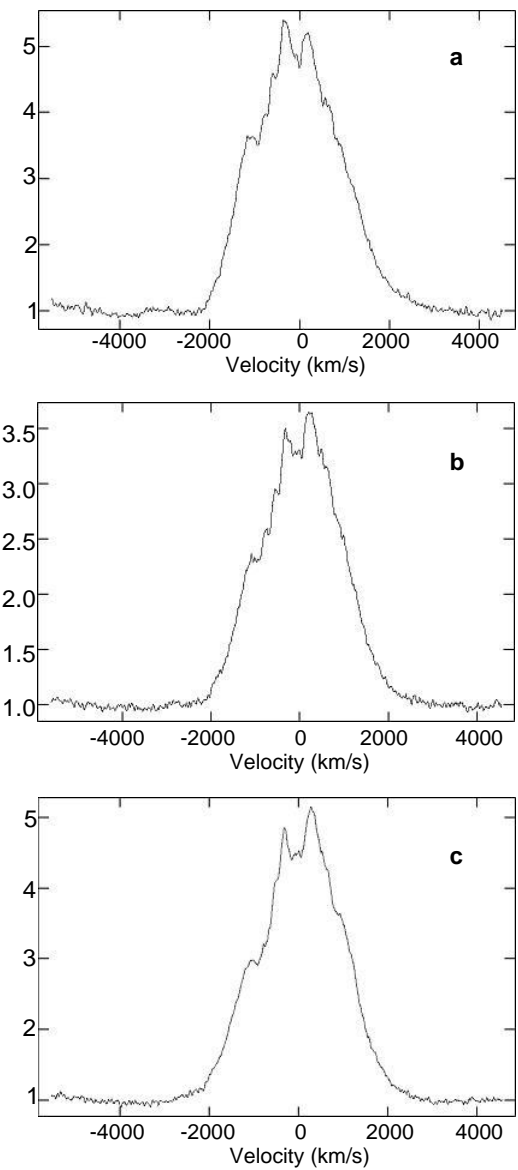

Fig. 1. Hα line profiles of V1065 CEN for **a:** 6 days (Jan31.9) **b:** 15 days (Feb 9.96) and **c:** 26 days (Feb 20.8) after maximum. The intensity is in relative flux.

The equivalent widths for the emission components of two profiles taken at 10 minutes interval were estimated using IRAF which are 38 °A and 40 °A respectively. The ultimate velocity of the expanding gas of the system can be determined by taking the red shift at the continuum level which gives around 2600 km/s.

## 3. PHOTOMETRIC ANALYSIS

### 3.1 Light Curve Analysis of V1065 CEN

The photometric analysis was carried out using the light curve data provided by AAVSO data archive. The Visual, V and B band light curves were plotted from 75 data points using MATLAB curve fitting tool. Since the V and B bands do not show peak magnitude, the visual light curve is used to find the maximum apparent magnitude and the corresponding time $t_o$ (2454126.486 JD = 2007 Jan. 25.98).

The maximum apparent magnitudes $m_v$ and $m_B$ are calculated from $m_v$-t and $m_B$-t graphs taking $t_o$ as 2454126.486 JD. In Fig. 3, Three arrows indicate our spectroscopic observations and the gradients were calculated for those three Julian days in order to explain the variations of Hα profiles. The rate of decline of the V-t light curve at $t_2$ is calculated as 0.075 and that for $t_3$ is 0.13.

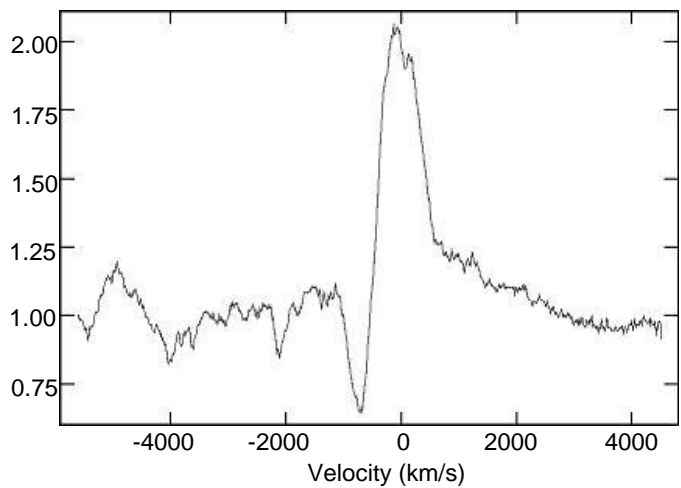

Fig. 2. Hα line profile of V1280 SCO observed at 4.14 days (Feb 20.9). The intensity is in relative flux.

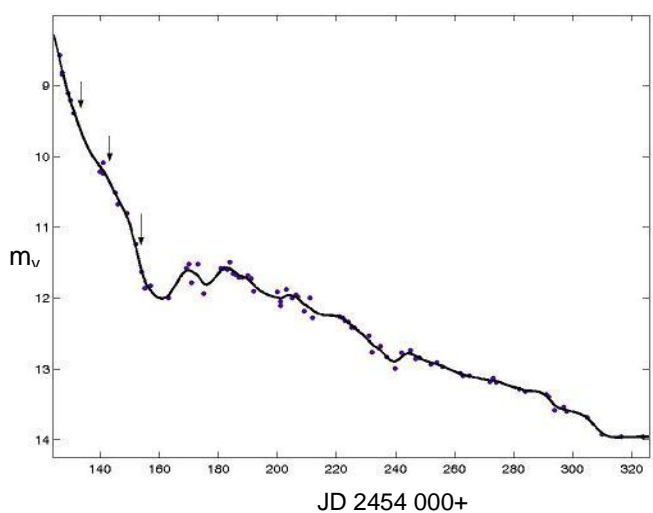

Fig. 3. V magnitudes vs Julian day for V1065 CEN. The arrows indicate the spectroscopic observations made on 6 days, 15 days and 26 days after maximum.

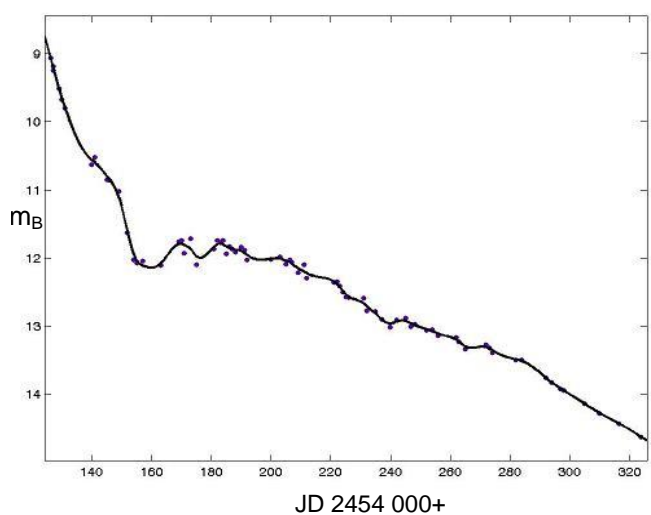

Fig. 4. B magnitudes vs Julian day for V1065 CEN

### 3.2 Light Curve Analysis of V1280 SCO

The light curve data for V1280 SCO are available only in visual band. The availability of 425 visual estimates provides to make 0.1 day bins taking the mean value of the individual points and 179 visual validated data points were used to plot the light curve. The best fit line is smoothing spline function with RMSE (0.2399) which is comparatively higher than that of V1065 CEN.

Due to the unavailability of V and B magnitudes the interstellar reddening and the visual extinction correction cannot be applied for V1280 SCO. The distance estimation was carried out by taking the time $t_2$ and $t_3$ from visual light curve. The arrow indicates the observation points of the Hα profile. The rate of decline at $t_2$ is 0.48 and for $t_3$ is 0.561.

Table 1. Basic Parameters of Nova V1065 CEN

| Parameter | V | B |
|---|---|---|
| Maximum apparent magnitudes | $m_V$ = +8.69±0.08 | $m_B$ = + 9.12±0.07 |
| Color index at maximum light | (B-V) = +0.43 ||
| Apparent magnitudes 15 days after maximum | $m_{V15}$ = +10.23 | $m_{B15}$ = +10.62 |
| Time taken to decline 2 magnitudes (t2) | $t_{2V}$ = 20.654 days | $t_{2B}$ = 22.604 days |
| Time taken to decline 3 magnitudes (t3) | $t_{3V}$ = 28.026 days | $t_{3B}$ = 30.885 days |
| Mean absolute magnitude at maximum | $M_v$ = -7.58±0.18 | $M_B$ = -7.75±0.25 |
| Intrinsic color index at maximum light | $(B-V)_{max} = M_B-M_v$ = -0.17 ||
| Distance modules | $m_V-M_v$=(8.69-(-7.58))=16.27±0.20 | $m_B-M_B$=(9.12-(-7.75))=16.87±0.26 |
| Mean distance modules | 16.57±0.33 ||
| Color excess at maximum light | $E_{B-V}$ = 0.43-(-0.17) = + 0.6 ||
| Visual extinction | $A_v = 3.2 \times E_{B-V}$ = 1.92 ||
| Unreddened distance module | 16.57-1.92=14.65 ||
| Distance to the nova | 8.51±0.33 kpc ||

## 4. DISCUSSION

### 4.1 Spectrometric Analysis

Very broad emission lines were observed (v=2000 km/s) in three spectral profiles of V1065 CEN and broadness of the emission lines were not considerably change with the time indicating the velocity gradient does not change much even though the magnitude drops by 3 from the maximum magnitude after 28 days.

Table 2. Basic Parameters of Nova V1280 SCO

| Parameter | Value |
|---|---|
| Maximum visual magnitude ($m_v$) | 4.0 ± 0.2 |
| Time at max. visual magnitude ($t_o$) | 2454148.25 |
| $t_2$ | 2454160.46 = 12.21 days |
| $t_3$ | 2454162.39 = 14.14 days |
| JD after 15 days from maximum | 2454163.25 |
| Apparent magnitude after 15 days $m_{v15}$ | 7.5 ± 0.2 |
| Mean absolute magnitude at maximum ($M_v$) | -8.7±0.1 |
| Distance module | $m_v - M_v$ = 12.7±0.2 |
| Distance to the nova | 3.2 ± 0.2 kpc |

According to the spectral classification [3], V1065 CEN is He/N-type spectra which characterize a broad (Gaussian FWHM 49 $^o$A), saddle shaped asymmetric Hα profile without absorption components. The Hα profile of V1280 SCO is completely different with V1065 CEN, shows prominent P-Cyg absorption and narrow emission line (Gaussian FWHM 26 $^o$A) which can be classified as Fe II type nova.

The early spectrum of V1065 CEN, Fig. 1(a) highly structured with few emission spikes due to the irregular density distribution in the gas expansion, declines to a smooth spectrum Fig. 1(c) which the ejecta turn to a more uniform density distribution with the expansion. The absence of emission spikes in V1280 SCO shows that the expansion is uniform than that of V1065 CEN and the density of the gas cloud is much more uniform.

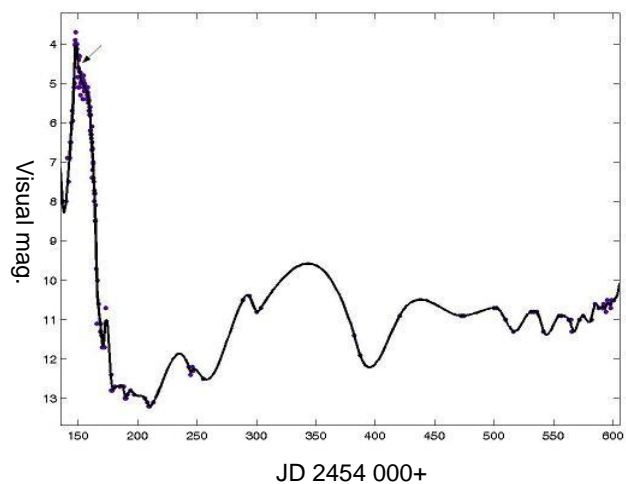

Fig. 5. Visual Magnitudes vs Julian Day for V1280 SCO

The P-Cyg profile is formed by the part of the wind directly between the observer and the nova. For an optically thin case if the source function of the wind is less than that of the background star, an absorption line will be appeared on the background intensity. If the source function of the wind is grater an emission line appears on the background intensity. There is a prominent P-Cyg profile can be observed in V1280 SCO suggesting that it was caused by a wind with a small source function. There is a possible very weak P-Cyg structure can be identified in the Fig. 1(a) of the V1065 CEN at around -2300 km/s and decline fast since there is no sign of P-Cyg in other two profiles. The absence of prominent P-Cyg structure suggests that the emission comes from discrete shell rather than a wind.

The presence of P-Cyg structure in the novae systems is highly depending on the line of sight and the equatorial bulge. The equatorial bulge may be caused around the white dwarf when the matter transfers from the red giant. The inclination of the line of sight to the plane of the bulge is a major factor for the appearance of profile and therefore the weak P-Cyg structure in V1065 CEN reveals the line of sight and the equatorial bulge is apart from some extent.

The expansion velocities of these two systems measured from the minima of the P-Cyg profiles are close to 2300 km/s for V1065 CEN 6 days after maximum and 716 km/s for V1280 SCO 4 days after maximum. Based on the spectroscopic observation the expansion of V1065 CEN is much higher than that of V1280 SCO and the Nova 1065 CEN is a fast nova compared to V1280 SCO.

Initially the blue component is much higher than red component of V1065 CEN and later profiles this was reversed. This behavior was observed in past observations of Hα profiles as well [4][5]. The possible explanation is the gas cloud is very dens at the beginning which obscure the material on the far side. When the gas and dust is thin enough the red component is immerging over the absorbed blue component.

The equivalent width measured for V1065 CEN on Jan. 26.3 (few hours before maximum – Jan. 25.98) was 930 $^{o}$A [1]. The equivalent width measured by us on Jan. 31.9, Feb. 9.96 and Feb. 20.8 are 210 °A, 120 °A and 190 °A respectively. There is a slight drop of equivalent width on the day 15, Fig. 1b and again increases on day 26, Fig. 1c. This variation is clearly matched with the light curve analysis. In the light curve, Fig. 3, three arrows indicate our spectral observations. The gradient of the light curve drops from 0.11396 to 0.0645 which corresponds to the intensity drop from plate **a** to plate **b**. Again the gradient increases to 0.1722 in the light curve explaining the raise in intensity in plate **c**. This is clearly indicates that the expansion is not uniform due to the radiation shells.

After 26 days of maximum the Hα profile of V1065 CEN, two emission peaks appeared on both sides of Hα(6563) and identified as NII(6584) and OII(6539). At the early stage of the nova the gas cloud is thick enough which the forbidden transitions are unlike. There are no nebular lines appeared in the vicinity of Hα but after 26 days the NII line became prominent reveals the nova become the nebular stage [6].

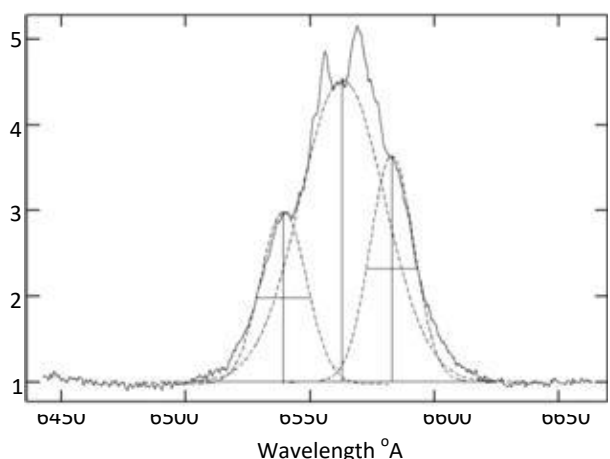

Fig. 6 Hα profile of V1065 CEN after 26 days of maximum. Blending Hα profile with NII(6583) and OII(6539).

### 4.2 Photometric Analysis

According to the classification system for nova light curves [7], the light cove of V1065 CEN is approximated to type $B_a$ which describes decline with minor irregular fluctuations with average $t_3$ is 35 days. The same calcification system categorizes nova V1280 SCO as type A which describes smooth fast decline without major disturbances and average $t_3$ is 12 days. The parameters calculated from the light curves $t_2$ and the rate of decline for V1065 CEN are 20 days and 0.08 respectively and for V1280 SCO are 12 days and 0.48 respectively. Based on these results the speed class of V1065 CEN is "fast" and V1280 SCO is "very fast" [8].

The availability of the light curve data for V and B bands for V1065 CEN allows to calculate $t_2$ and $t_3$ separately for V and B bands and hence the absolute magnitudes, $M_V$, $M_B$ were calculated using the MMRD relationships [9]. The parameters $t_{2V}$ and $t_{3V}$ are used in the MMRD relations: [1][3] and the relations 15 days after maximum: [1] and obtained the mean absolute magnitude, $M_V$ (table 1). The parameters $t_{2B}$ and $t_{3B}$ are used in the MMRD relations: [8][6] and the relation 15 days after maximum: [8] and obtained the mean absolute magnitude, $M_B$ (table 1). Hence the intrinsic color index ($M_B-M_V$) at maximum light is calculated as -0.17 and the color index ($m_B-m_V$) at maximum light taken from the light curves B and V is +0.43. The results show the interstellar reddening is highly affected to the distance module and the visual extinction is 13% of the unreddened distance module. The distance for nova V1065 CEN is 8.51 ± 0.33 kpc, relatively higher value which causes higher visual extinction. The derived value for the distance of nova V1065 CEN, 8.51 ± 0.33 kpc, is in good agreement with the estimate of the distance: Andrew Helton (University of Minnesota) 7.9 kpc.

The visual extinction is not corrected for the distance estimation of V1280 SCO due to the unavailability of light curve data in the B and V bands. The available visual light curve data were used to plot the light curve (Fig. 5) and hence deduced $t_2$ and $t_3$ and substitute the MMRD relationships: [1] and the relations 15 days after maximum: [1] to calculate the mean absolute magnitude, $M_V$ (Table 2) and the estimated distance is 3.2 ± 0.2 kpc.

## 5. CONCLUSIONS

The nova V1065 CEN is originated by He/N type nova based on the observation of broad emission line in the H$\alpha$ and narrow emission lines which reveals nova V1280 SCO is Fe II type. The photometric analysis of the two systems convey the system V1065 CEN is a comparatively slower than the system V1280 SCO. The estimated distance with the correction of interstellar extinction for the nova V1065 CEN is 8.51 ± 0.33 kpc and this is for V1280 SCO is 3.2 ± 0.2 kpc without the correction of the interstellar reddening. The prominence of the P-Cyg structure is highly depending on the different geometry of the binary system to the line of sight of the observer. The coincidence of the line of sight to the equatorial bulge causes the prominent P-Cyg structure in the H$\alpha$ profile in V1280 SCO while the deviation of this geometry reveals a very weak P-Cyg profile in V1065 CEN.

**ACKNOWLEDGEMENT**

**References**
[1] W. Liller, Vina del Mar, 2007, IAU Circ. No. 8800
[2] H. Yamaoka, 2007, IAU Circ. No. 8803
[3] R. E. Williams 1992, AJ, 725, 733
[4] L. L. Kiss, J. R. Thomson, 2000, A&A, L9, L12
[5] S. E Smith., P. V. Noah, M. J. Cottrell, 1979, Astronomical Society of the Pacific, 775, 780
[6] Xiaobin Zhang, Yulian Guo, 1992, IAU Circ. No. 3783
[7] H. W. Duerbeck, 1980, Astronomical Society of the Pacific, 165, 176
[8] M. F.Bode, A. Evans, 2008, Classical Novae, Cambridge University Press, p. 1-76, p. 195-230,
[9] D. Chochol, L. Hric, Z. Urban, et al., 1993, A&A, 103,113
[3] M. Della Valle, M. Livio, 1995, AJ, 704, 709
[8] W. Pfau, 1976, A&A, 113,115
[6] M. Livio, 1991, AJ, 516, 522